\documentclass[namedreferences]{spackap}

\usepackage{graphicx} % used for embedding postscript figures
\usepackage{mathptm} % used for postscript
\begin{document}

\begin{article}

\begin{opening}

\title{Current Challenges Facing Planet Transit Surveys}
\runningtitle{Current Challenges Facing Planet Transit Surveys}
%\subtitle{}

\author{D. \surname{Charbonneau}\email{dc@caltech.edu}}
\runningauthor{D. Charbonneau}
\institute{California Institute of Technology, 105-24 (Astronomy),\\
1200 E. California Blvd., Pasadena, CA 91125 USA}

\received{\ldots} \accepted{\ldots}

\begin{ao}
D. Charbonneau, California Institute of Technology,
105-24 (Astronomy), 1200 E. California Blvd., Pasadena, 
CA 91125 USA, dc@caltech.edu
\end{ao}

\begin{abstract}
The initial task that confronted extrasolar-planet transit surveys 
was to monitor enough stars with sufficient 
photometric precision and complete phase coverage.  
Numerous searches have been pursued over the
last few years.  Among these projects are shallow, 
intermediate, and deep surveys of 
the Galactic plane, and monitoring of open clusters, 
and a globular cluster.
These projects have all defeated the initial technical challenge, 
but a new obstacle has risen in its place:  Single-color photometric 
time series are not sufficient to identify uniquely 
transiting planet systems, as eclipsing binary stars can
mimic the signal.  Multicolor photometric 
time series and multi-epoch spectroscopic monitoring are required 
to cull the list of candidates prior to
high-precision radial velocity monitoring.  I also discuss the
prospects for detecting another transiting system among the
planets found by the radial-velocity method, as well as review
the recent announcement of \mbox{OGLE-TR-56~b}, the first extrasolar
planet detected by the transit method.
\end{abstract}

\end{opening}

\section{Introduction}
\label{sec:introduction}
Among the many great changes invoked by the \inlinecite{mayor95}
detection of the planet orbiting 51~Pegasi were those 
concerning the potential use of photometric transits to
detect and characterize extrasolar planets.
Prior to 51~Peg~b, several papers \cite{rosenblatt71, borucki84, borucki85, giampapa95}
had outlined the chief obstacle facing the transit method:
Ground-based photometry was likely to succeed only for gas-giant
planets, yet such planets were expected only at large distances.
Even if all Sun-like stars had a Jupiter, transiting systems would 
be rare (only 1 in 1000 systems would have the inclination 
to transit), and, worse yet, the transit event in such
systems would occur only once per 12-year orbital period.

In contrast to this scenario, there are now more than 20 active transit
surveys (see \opencite{horne03} for a complete listing\footnote{See also 
\texttt{http://star-www.st-and.ac.uk/$\sim$kdh1/transits/table.html}, 
maintained by K.~Horne}), nearly all of 
which are focussed on detecting analogs of 51~Peg
(which I shall refer to here as hot Jupiters).  
I remind the reader briefly of the characteristics
of the signal the transit searchers seek:  The amplitude of the flux
decrement is roughly $(R_{p}/R_{*})^{2} \simeq 0.01$,
where $R_{p}$ is the radius of the planet,
and $R_{*}$ is that of the star.
Transits occur once per 3-7~day orbital period,
and last 2-4~hours.  The rate of occurrence of hot Jupiters for Sun-like
stars is currently estimated at $r=0.0075$ \cite{butler01}, and the likelihood
of a hot Jupiter system with a semi-major axis $a$
presenting a transiting inclination is 
$p \simeq (R_{*}/a) \simeq 0.1$ 
(for a uniform distribution of orbital inclinations).  Assuming that
complete phase coverage is achieved, the number of
stars that must be examined to find one transiting hot Jupiter system
is $n = 1/(r \, p \, \, g) \simeq 1300/g$, where g is the fraction
of stars examined that are ``good'' targets, i.e. Sun-like and not 
members of a close binary.

\section{Photometry of Stars with Doppler-Detected Planets}
\label{sec:doppler}
Numerous groups have performed high-precision photometry
on the known planetary systems with the smallest
semi-major axes, and transits by gas-giant objects
have been ruled out (in order of increasing semi-major axis) for
HD~83443 (S.~Udry, personal communication),
HD~46375 \cite{henry00}, HD~179949 \cite{tinney01}, 
HD~187123 \cite{castellano00}, $\tau$~Boo \cite{baliunas97, henry00a}, 
BD-10{\deg}3166 \cite{butler00}, HD~75289 (S.~Udry, personal communication),
51~Peg \cite{henry97, henry00a}, $\upsilon$~And~b \cite{baliunas97, henry00a},
HD~49674 \cite{butler02}, HD~168746 \cite{pepe02}, HD~108147 \cite{pepe02} \&
55~Cnc~b \cite{baliunas97, henry00a}.

It is worthwhile to consider whether these non-detections are
consistent with our expectations.  If the probability
of a transiting configuration is $p$, then the probability
of finding $k$ transiting systems of $n$ stars examined is
$P = n!/[k! (n-k)!] (1-p)^{n-k} p^{k}$.  If we assumed that 
each of the 14 systems that were monitored (HD~209458 and the 13 listed 
above) had a probability of presenting transits of $p \simeq R_{*}/a \simeq
0.1$ (for the Sun, this corresponds to a semi-major axis of 0.047~AU),
then the chance of finding one (and only one) transiting system
is 0.36.  The chance of having found no such systems is 0.23.
On the other hand, the chances of having found 2 or 3 such systems
are 0.26 and 0.11 respectively.  Any of these outcomes
would have been roughly consistent with a uniform distribution
of orbital inclinations (and the scenario that materialized was
the single most likely one).

We needn't assume a uniform value of $p=0.1$ for all
systems; we can estimate $R_{*}$ from parallax measurements
(or stellar modeling), and $a$ is calculated from the radial velocity
period and an estimate of the stellar mass $M_{*}$.  
Excluding HD~209458, one can then ask what the probability
was of examining the other 13 systems and finding no transits.
This value is given by $P = \prod_{i=1}^{13} \; (1 - {R_{*,i}}/{a_{i}}) 
\simeq 0.26$ (where I have assumed values for $R_{*}$ gathered from
the literature).  This number is also consistent with a uniform distribution
of orbital inclinations.

It is reasonable to ask whether photometric monitoring of additional
systems is worthwhile.  Such observing campaigns are increasingly
difficult for longer periods, as the transits are more infrequent,
and the uncertainties in the predicted times of the events are
greater.  With these considerations in mind, G.~Laughlin has
established a project\footnote{See \texttt{http://www.transitsearch.org}, 
maintained by G. Laughlin.} to motivate amateur astronomers
to pursue the most promising of the remaining extrasolar planet
systems.  The requisite photometric precision can be achieved
with amateur-grade, commercially-available CCD cameras.  Amateur 
telescopes provide more than enough aperture
to gather the requisite flux (recall that \opencite{charbonneau00}
used a 10~cm aperture Schmidt camera to record the first transits
of HD~209458), although the paucity of sufficiently bright calibrator
stars in the typical field-of-view of such instruments can be a problem.  
This network has recently claimed to rule out transits
for HD~217107 \cite{fischer99}.  There are 24 extrasolar planets 
with periods less than 200~days
that have not yet been examined for transits.  For this sample,
the probability of at least one transiting system is  
$P = 1 - \prod_{i=1}^{24} \; (1 - {R_{*,i}}/{a_{i}}) \simeq 0.62$
(using values for $R_{*}$ as listed at the project website).

Photometric monitoring of Doppler-detected planet systems is 
also a useful check
that the radial velocity variations are due to an orbiting planet,
and not intrinsic stellar variability.  Recently, \inlinecite{henry02}
presented photometry of HD192263 showing variability at the 
RV period, and casting the planet interpretation \cite{santos00, vogt00} 
into doubt.

Although there may be a few more transiting planets in the
current radial velocity sample, the ongoing Doppler surveys
will not provide a substantial population of such objects.  The
primary goal of these surveys is to characterize the planet population
at large semi-major axes.  As a result, these surveys will 
continue to monitor the current target list (comprising some 1500 stars)
for many years to come, but will not add many new targets.  Since hot 
Jupiters are the most quickly and easily detected, the current Doppler precision
of $3 {\rm \ m \, s^{-1}}$ has likely revealed the majority of such objects
with masses greater than $0.2~M_{\rm Jup}$ in the current target list.
The desire for a large number of transiting hot Jupiters, and
the realization that the current Doppler surveys are unlikely
to provide this sample, motivates the various transit searches
that are the subject of the remainder of this paper. 

\section{Radial Velocity Follow-Up of Transit Candidates}
\label{sec:basics}
Transiting extrasolar planets are of substantial value
only if the radial velocity orbit can be measured.
Since the current transit surveys target stars ranging in brightness
from $9 \le V \le 21$, it is worthwhile to consider the
resources that will be required to accomplish this follow-up
measurement once candidate systems have been identified.

For typical Sun-like stars with rotational velocities at or below
a spectrograph resolution $R$, the Doppler precision may be roughly 
estimated by \cite{brown90}:
\begin{equation}
\delta v_{rms} \simeq \frac{c}{R \ d \ (N_{lines} \ N_{pix} \ I_{c})^{1/2}},
\label{vrms}
\end{equation}
where $c$ is the speed of light, $d$ is the typical fractional line depth (relative 
to the continuum), $I_{c}$ is the continuum intensity per pixel, 
$N_{lines}$ is the number
of spectral lines, and $N_{pix}$ is the number of pixels per line.
As an example, consider the precision that may be expected from
the HIRES spectrograph on the Keck~I telescope: 
Assuming $d=0.4$, $N_{pix} = 2$, and $N_{lines} = 100$, and a resolution
$R = 70~000$ and count level $I_c = 90~000$ (corresponding to a 5~minute
exposure of a late G star at $V=8$), then the formula above yields 
\mbox{$\delta v_{rms} = 2.5 {\rm \ m \, s^{-1}}$}.    G.~Marcy 
(personal communication) reports typical photon-limited 
Doppler errors of \mbox{$3 {\rm \ m \, s^{-1}}$} for 5~minute integrations 
on V=8 G-dwarf stars, in keeping with this estimate.
(In practice, the achieved Doppler precision results from the use of more 
lines than that assumed here,  at a variety of line depths and
signal-to-noise ratios).

For radial-velocity follow-up of stars with candidate transiting planets, 
we are interested to know how much telescope
time $t_{obs}$ will be required to detect (or exclude) a secondary mass $M_{p}$ orbiting
a star of mass $M_{*}$ with a period $P$.  To derive this relation,
I start with amplitude of the radial velocity signature induced on
the primary, 
\begin{equation}
K_{*} = \left( \frac{2 \ \pi \ G}{P} \right)^{\frac{1}{3}} {M_{p} \sin{i}} \ \, {M_{*}^{-\frac{2}{3}}}.
\end{equation}

Equation \ref{vrms}, and the experience of the radial velocity observers,
tells us that a precision of $3 {\rm \ m \, s^{-1}}$ is obtained with Keck/HIRES
in 5 minutes on a $V=8$ star.  I assume that the Doppler precision is 
photon-noise-limited, and that the amplitude must exceed 
4 times the precision to be secure (in keeping with the rule-of-thumb 
suggested by \opencite{marcy00}).  Equating these requirements,
\begin{equation}
\frac{K_{*}}{4} = \delta v_{rms} = 3 {\rm \ m \, s^{-1}} \left[ \left( \frac{t_{obs}}{5~{\rm min}} \right) 10^{-\frac{V-8}{2.5}} \right]^{-\frac{1}{2}},
\end{equation}
I solve for the required integration time,
\begin{equation}
t_{obs} = 0.0363~{\rm min} \left( \frac{M_p}{M_{\rm Jup}} \right)^{-2} \left( \frac{M_{*}}{M_{\sun}} \right)^{\frac{4}{3}} \left( \frac{P}{3~{\rm days}}\right)^{\frac{2}{3}} 10^{\frac{V-8}{2.5}}. 
\label{tobs}
\end{equation}

Equation~\ref{tobs} makes it clear that the required integration time
is very sensitive to the planetary mass and the stellar brightness (as
one would expect).  It is important to avoid the temptation
to assume $M_{p} = 1~M_{\rm Jup}$ in calculating $t_{obs}$.  
Of the 17 planets with $a \le 0.1$~AU, 70\% have masses below
1~$M_{\rm Jup}$, and the median mass is $0.5~M_{\rm Jup}$.
Thus if one adopts the value of $r=0075$ (see \S\ref{sec:introduction})
in planning a transit 
search, one should also adopt a mass of 
$0.2~M_{\rm Jup}$ in calculating the required $t_{obs}$ 
with equation~\ref{tobs}.  The result is to increase
the predicted integration times by a factor of 25.

\section{A Selection of Transit Surveys}
\label{sec:surveys}
While it is not possible here to review all current transit
searches, I have selected four surveys which 
straddle the range of current efforts.  I briefly
review the status of the projects, before summarizing
the prospects for Doppler (and additional) follow-up observations
of candidates yet-to-be identified by these surveys.

\subsection{Shallow, Wide-Angle Surveys}
\begin{figure}
\centerline{
\includegraphics[height=0.4\textheight]{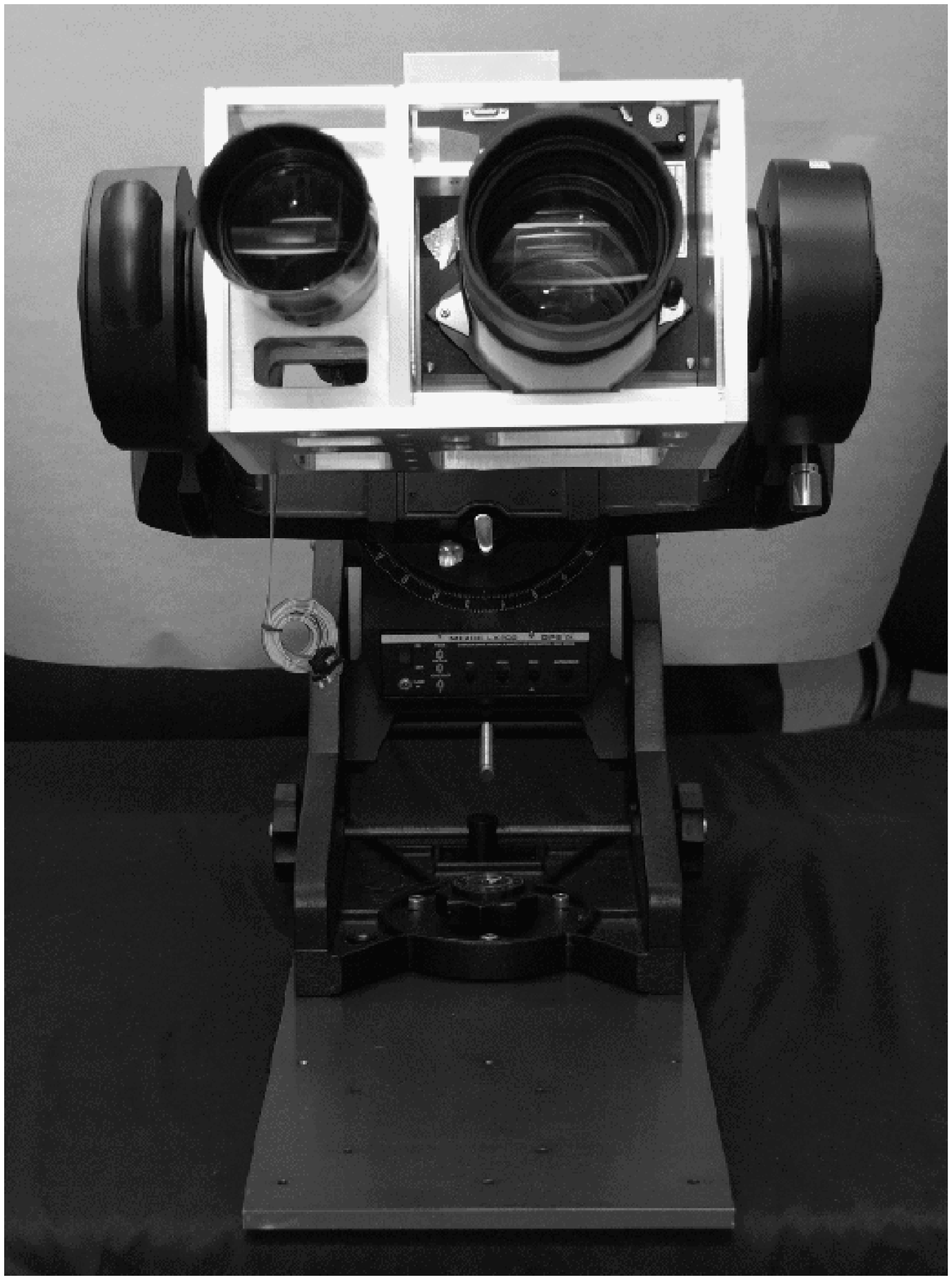}\qquad
\includegraphics[height=0.4\textheight]{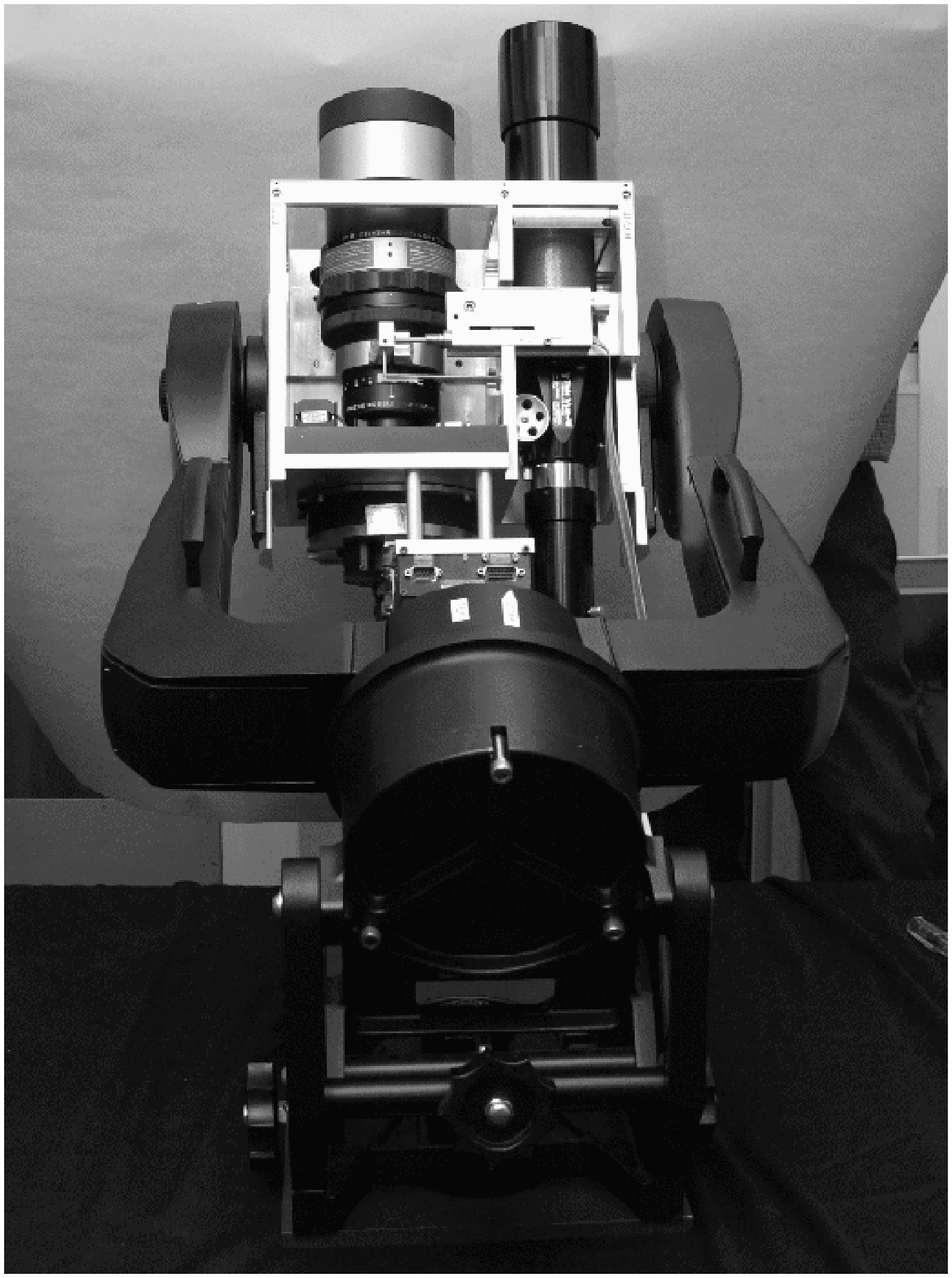}
}
\caption{South-facing (left panel) and north-facing (right panel) views
of the Palomar planet search instrument.  The primary imaging lens
is on the left side of the camera as shown in the right panel;
a 280mm f/2.8 commercial lens images a $5.7{\deg}{\times}5.7{\deg}$
patch of the sky onto an 2k${\times}$2k thinned CCD.  The system
is assembled almost entirely from amateur-grade, commercially-available
parts, with the exception of the primary CCD camera.}
\label{camera1}
\end{figure}

Working with J.~T.~Trauger's team at NASA/JPL, I have assembled
a small-aperture wide-field transit-search instrument (Figure~\ref{camera1}) 
that typifies 
many extant systems, such as STARE \cite{brown00} and 
Vulcan \cite{borucki01}.  The system
(with the exception of the primary CCD camera) is built entirely
of commercially-available items typically intended for
amateur use.  I describe the basic features here for the benefit
of those readers contemplating the fabrication of a similar system.
The main optic is an f/2.8 280mm camera lens imaging a 
$5.7{\deg}{\times}5.7{\deg}$
patch of the sky onto an 2k${\times}$2k thinned CCD.
A micrometer allows for automated focus adjustments.  Each CCD 
pixel is 13.5${\mu}$m, corresponding to 10~arcsec.  A filter
wheel houses the  SDSS g', r', i', z' and Bessell R filters.
For guiding, an f/6.3 440mm lens feeds a commercial 
prepackaged CCD guide camera.  The 
system is mounted in an equatorial fork mount,
and housed in a refurbished clamshell enclosure at Mt. Palomar
in southern California.  All systems are operated by a single 
Linux-based workstation, and a command script guides 
the instrument through each night's observing.

This system is the third in a network:  The other two instruments are
STARE (PI: T.~M. Brown), located in the Canary Islands, and 
PSST (PI: E.~W. Dunham), located in northern Arizona.  
Each telescope produces a time series of R-band images (with
typical integration times of 2 minutes), and only one field is monitored
for a typical observing campaign of 2~months.  
We perform weighted-aperture photometry
on these images to produce a photometric time series for each star.
(Image subtraction methods, such as those as described by \opencite{alard98},
are unlikely to result in a 
significant increase in precision because the 10~arcsec pixels
yield slightly undersampled and seeing-independent images).
In a typical field-of-view centered on the Galactic plane,
roughly 6000 stars ($9 \le V \le 11$) are monitored with 
sufficient accuracy to detect 
periodic transit-like events with an amplitude of 1\%.

\begin{figure}
\centerline{
\includegraphics[width=0.95\textwidth]{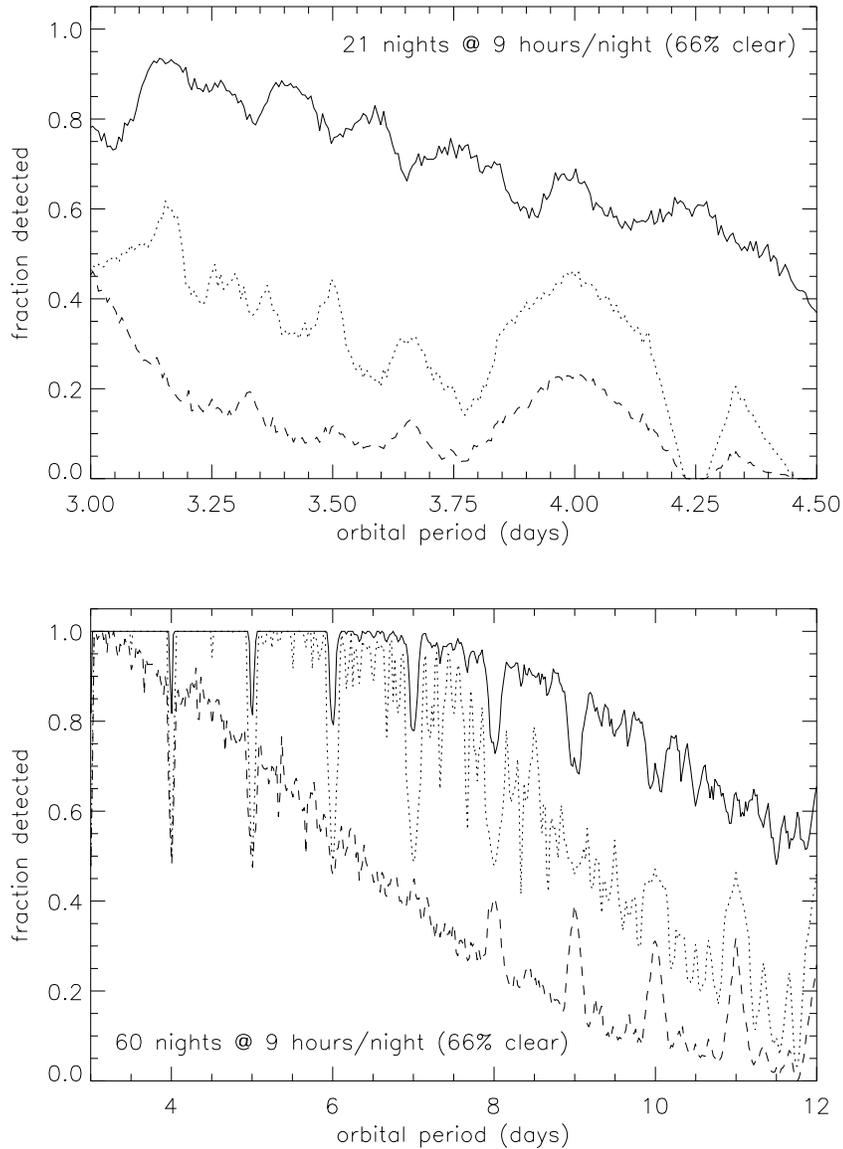}
}
\caption{The upper panel shows the recovery rate 
for a 3-week campaign to find transiting planets
(assuming 9 hours per night, and 66\% clear weather), 
as a function of orbital period.  
Three half transits are required.  
The dashed line is the result for a single telescope, the
dotted line is that for three telescopes at the same longitude, and
the solid line is that for a 3-element network with a telescope 
in each of the Canary Islands, Arizona, and California.  The lower
panel shows the corresponding recovery rates for a 2-month campaign
with the same night length and weather.
In both panels, the network has nearly exhausted the 
field (and hence a new field can be monitored, increasing the total
number of targets), whereas both the
single-element and single-longitude systems are still lacking
significant phase coverage.}
\label{network}
\end{figure}

For a single telescope, the primary losses in efficiency are 
due to the day-night cycle, and weather.  Currently the telescopes are 
operated as stand-alone systems monitoring the same field-of-view.
Once networked so that the time series are combined prior to 
performing the search for transit-like signals, we expect a 
substantial increase in efficiency.
Specifically, we anticipate that the 2~months required 
by a single instrument to achieve 85\% completion
(for orbital periods less than 4.5~days) 
can be reduced to only 3~weeks (Figure~\ref{network})
with the current longitudes afforded by the three sites.
Since fields are exhausted much more quickly, the number of
stars monitored in a year of operation is greatly increased.

\subsection{Intermediate Galactic Plane Surveys -- OGLE~III}
The team operating the Optical Gravitational Lensing Experiment has 
conducted a search for low-amplitude transits in three
fields in the direction of the Galactic center \cite{udalski02, udalski02a}.  
They obtained 800 $I$-band epochs per field spanning 32 nights.  
More than 5 million stars were monitored, to which
they applied a substantial cut in color-magnitude 
space to reduce the number of targets to 52~000 disk stars ($14 \le V \le 18$)
with photometry better than 1.5\%.  Of these, 59 candidates were 
identified with flat-bottomed eclipses with depths less than 8\%.

\inlinecite{dreizler02} presented spectroscopy of some 
of these candidate stars, with the goal of obtained secure
spectral classifications.  The stellar radii they inferred
allowed them to calculate more accurate radii of the transiting
objects.  In many cases, the \mbox{OGLE-III} objects had radii that
were too large for planet-mass bodies.

After a similar (but independent) spectroscopic reconnaissance, \inlinecite{konacki03}
found that 6 of the candidates had solar-type spectra with
no radial-velocity variation at the level of a few ${\rm km \, s^{-1}}$.
Subsequent radial velocity monitoring with Keck/HIRES revealed
that one candidate, \mbox{OGLE-TR-56}, showed a velocity change with
an amplitude of $167{\rm \ m \, s^{-1}}$ and consistent with the
1.2-day photometric variation.  The team performed numerous modeling tests 
to rule out spectral blends (a significant concern, as the
\mbox{OGLE-III} fields are very crowded), 
leading them to announce the
first detection of an extrasolar planet by the transit method.
The new-found planet has a mass $M_{p} = 0.9 \pm 0.3 \ M_{\rm Jup}$
and radius $R_{p} = 1.3 \pm 0.15 \ R_{\rm Jup}$, and hence a
density of $\rho = 0.5 \pm 0.3 \ {\rm g \, cm^{-3}}$, similar
to the more precise estimates for HD~209458~b \cite{brown01}.  The current
radial-velocity phase coverage is sparse, but will likely be filled
in during the 2003 bulge season.  The most surprising result
is the 1.2-day orbital period.  The Doppler surveys have
found no orbital periods below 2.99~days, and a large
pile-up of objects at this value (8 planets with periods less than 4 days).

\subsection{Deep Galactic Plane Surveys -- EXPLORE}
The deepest transit search currently underway is the EXPLORE project
\cite{mallen02, yee02}.  The EXPLORE team conducts observing campaigns
(lasting typically 2 weeks) 
using the CTIO 4-m and CFHT 3.6-m telescopes to continuously
monitor fields in the Galactic plane in $I$-band.  EXPLORE-I (southern
hemisphere) received 11 nights in 2001 and 
delivered a photometric precision of better than 1\%
on 37~000 stars with $14.5 \le I \le 18.2$.  EXPLORE-II (northern
hemisphere) received 14 nights in 2001 and delivered a similar photometric
precision of better than 1\% on 9500 stars.  Both surveys
acquired data with a similar sampling interval of 2.7~minutes.
The team has identified several candidates, and spectroscopic follow-up has
been conducted on VLT/UVES and Keck/HIRES.

The primary benefit of the EXPLORE campaign is that they will probe
a significant number of low-mass (K \& M) main sequence stars.
This is in contrast to the wide-field surveys, which will probe mostly F \& G 
stars (as less massive stars are not a significant fraction of
the $V < 11$ population).  The principal challenge facing EXPLORE
is that even preliminary follow-up radial velocity monitoring (i.e.
with the goal of ruling out eclipsing binary star systems) is very resource 
intensive, requiring 8-m class telescopes.

\subsection{Open Cluster Surveys -- PISCES}
The identification of open clusters as good targets for
transit searches was pointed out by \inlinecite{janes96}.
Open clusters offer ideal laboratories in which to study 
the characteristics of hot Jupiters, since
the stars share a common metallicity and age.  The search
for transiting planets in open clusters is motivated
in part by the conclusion by \inlinecite{gilliland00}
that the population of hot Jupiters in the globular
cluster 47~Tuc is greatly depleted relative to the
local solar neighborhood.  Two of the many contending explanations
for this result are (1) the low-metallicity environment
results in a reduced formation and/or migration rate of
Jupiter-mass planets, or (2) protoplanetary
disks are tidally disrupted by the close passage of
stars in the crowded environment of a globular cluster.  
Open clusters offer
environments at a range of metallicities, yet are
not crowded enough for disruption by stellar
encounters to be significant.  There is an
additional benefit for searching for transiting planets
in open clusters: interpretation of candidates is greatly simplified,
since the stellar radius and mass can be reliably assumed
from the cluster color-magnitude diagram.  

\inlinecite{mochejska02} present results of a month-long 
campaign on the open cluster NCG~6791.  They demonstrate
adequate photometric precision for the detection 
of hot Jupiters, and present lightcurves for 62 variable
stars (but no planet candidates).  The challenges
facing their survey are representative of those for
other open-cluster searches (e.g. \opencite{street02}):
Open clusters typically contain several thousand stars, and
thus would yield only a handful of detections even if
complete phase coverage can be obtained.
Specifically for \inlinecite{mochejska02}, NGC~6791
has roughly 10~000 member stars, of which 4110 (59\%) and 
2053 (29\%) have sufficient photometric precision to
detect transits for planets with radii of 1.5 and 1.0~$R_{\rm Jup}$, 
respectively.  Furthermore,
these stars are very faint ($17 \le R \le 21$), and thus 
even preliminary follow-up spectroscopy 
(i.e. with the goal of ruling out eclipsing binary
star systems) will likely require 8-m class telescopes.

\section{The Task at Hand}
Several years ago, the principal technical challenge facing
proposed transit surveys could be summarized in the following
question:  Could enough stars ($\sim$10~000) be surveyed with sufficient
precision ($\sim3$mmag) and for enough nights to obtain complete phase coverage?
The good news is that the answer to this question is a 
definite yes, as evidenced by the diversity of projects
described in \S\ref{sec:surveys}.  However, a new challenge has arisen,
which I will broadly describe as sorting out the ``false positives''.
Single-color photometric time series alone are not
sufficient to identify uniquely the transiting-planet systems.
There are at least three general forms of apparent
variability that can mimic these signals, all of which
involve an eclipsing binary star system.  The first
is grazing incidence equal-size stellar binaries,
such that the occulted area is roughly 1\% (and the
period is underestimated by a factor of 2).  In general,
this contaminant can be ruled out by sufficiently precise
and rapid photometric observations during transit,
as grazing incidence binaries will present a V-shaped
(as opposed to a flat-bottomed) eclipse.  
The second contaminant is a central transit by
a smaller star in front of a larger one.  Multi-epoch
low-precision (\mbox{$\sim {\rm \ km \, s^{-1}}$}) 
radial-velocity monitoring would identify such systems,
as the amplitude of the radial velocity orbit would
be orders of magnitude larger than that expected for
a hot Jupiter.  
The third and most insidious contaminant is a stellar blend, 
where an eclipsing binary (with central transits) has its eclipse 
depth diluted down to $\sim$1\% by the light
of a third star (either physically associated, or
simply lying along the line of sight).  In general,
multi-color photometry of such a system should
reveal a color-dependent transit depth, whereas 
the transit depth for a hot-Jupiter system
should vary only slightly (due to the effects of stellar
limb-darkening).  For examples of such systems, 
and follow-up spectroscopy revealing
their true (non-planetary) nature, see \inlinecite{borucki01}.

In summary, most contaminating systems can be rejected by
either (1) rapid-cadence, multi-color photometry of
the transit curve revealing either a V-shaped eclipse,
or a large color-dependence to the transit depth, or 
(2) multi-epoch spectroscopy revealing large Doppler
variations, consistent with a massive secondary.

Transiting planets are of significant value
only if both their mass and radius can be estimated.
Moreover, the reality of putative planets discovered by transit
photometry but without a measured radial-velocity
orbit may be doubted.  Conversely, the 
detection of a radial-velocity orbit at the same
period and phase as
the ones derived from photometry is strong
evidence supporting the planet interpretation.

As an illustration of the dramatic effects of target brightness
on the prospects for detecting the radial velocity orbit,
consider Keck/HIRES observations of 
stars with $V=10.5$ (typical brightness for
a target star in the wide-angle surveys), $V=15.3$ (the
brightness of \mbox{OGLE-TR-56}) and $V=18$ (faint star
in a cluster survey or deep Galactic plane search). 
Using equation~\ref{tobs}, at $V=10.5$, a detection of
a $0.2~M_{\rm Jup}$ planet can be achieved with
integrations of only 10~minutes.
For \mbox{OGLE-TR-56} ($M_{p} = 0.9~M_{\rm Jup}$, $P = 1.2$~d),
integrations of 20~minutes are required (indeed, this
is close to what \opencite{konacki03} used).  However, less massive planets
at longer orbital periods ($M_{p} = 0.5~M_{\rm Jup}$, $P = 3$~d)
are a challenge, requiring 2~hours of integration
per measurement.  At $V=18$, the situation is very difficult
indeed:  A 1-hour integration yields a detection
limit of only $2.5~M_{\rm Jup}$ for a 3-day orbital period.
A planet with a mass of $1~M_{\rm Jup}$ would require more
than 6~hours of integration per observation (and recall
than most of the known hot Jupiters have masses below
$1~M_{\rm Jup}$).  In summary, radial velocity measurements
are straightforward for the typical targets in the wide-angle surveys,
challenging but feasible for the intermediate galactic plane
surveys (targets brighter than $V=16$), and unlikely to
succeed for targets toward the faint end of the deep
Galactic plane searches, and some open cluster surveys.  Finally, 
it is important to note another strong reason to favor
bright stars:  Many follow-up measurements of HD~209458 are now being
vigorously pursued (see \opencite{charbonneau03} for a summary),
and most of these are photon-noise limited.
While some of these techniques may be feasible for candidates emerging from
the wide-angle surveys (with stars typically 10 times fainter
than HD~209458), they are unlikely to approach a useful precision
for fainter stars.
%

%\acknowledgements 

\end{article}


\begin{thebibliography}{}

\bibitem[\protect\citeauthoryear{Alard and 
Lupton}{1998}]{alard98} Alard, C.~and Lupton, R.~H.:  1998,  'A 
Method for Optimal Image Subtraction', {\it Astrophysical Journal} 
{\bf 503}, pp.~325

\bibitem[\protect\citeauthoryear{Baliunas et 
al.}{1997}]{baliunas97} Baliunas, S.~L., Henry, G.~W., Donahue, 
R.~A., Fekel, F.~C., and Soon, W.~H.:  1997,  'Properties of Sun-like Stars 
with Planets: ${\rho}^{1}$~Cancris, $\tau$~Bootis, and $\upsilon$~Andromedae', {\it 
Astrophysical Journal} {\bf 474}, pp.~L119

\bibitem[\protect\citeauthoryear{Borucki et 
al.}{2001}]{borucki01} Borucki, W.~J., Caldwell, D., Koch, D.~G., 
Webster, L.~D., Jenkins, J.~M., Ninkov, Z., and Showen, R.:  2001,  'The 
Vulcan Photometer: A Dedicated Photometer for Extrasolar Planet Searches', 
{\it Publications of the Astronomical Society of the Pacific} {\bf 
113}, pp.~439--451  

\bibitem[\protect\citeauthoryear{Borucki et 
al.}{1985}]{borucki85} Borucki, W.~J., Scargle, J.~D., and 
Hudson, H.~S.:  1985,  'Detectability of Extrasolar Planetary Transits', 
{\it Astrophysical Journal} {\bf 291}, pp.~852--854  

\bibitem[\protect\citeauthoryear{Borucki and 
Summers}{1984}]{borucki84} Borucki, W.~J.~and Summers, A.~L.:  
1984,  'The Photometric Method of Detecting Other Planetary Systems', {\it 
Icarus} {\bf 58}, pp.~121--134

\bibitem[\protect\citeauthoryear{Brown et al.}{2001}]{brown01} 
Brown, T.~M., Charbonneau, D., Gilliland, R.~L., Noyes, R.~W., and Burrows, 
A.:  2001,  'Hubble Space Telescope Time-Series Photometry of the 
Transiting Planet of HD 209458', {\it Astrophysical Journal} {\bf 
552}, pp.~699--709  

\bibitem[\protect\citeauthoryear{Brown and Charbonneau}{2000}]{brown00} 
Brown, T.~M.: 2000, 'The STARE Project: A Transit Search for Hot Jupiters',
in {\it ASP Conf. Ser. 219: Disks, Planetesimals, and Planets}, 
F. Garzon, C. Eiroa, D. de Winter, and T. J. Mahoney (eds.), ASP,
San~Franciso, pp.~584--589, astro-ph/0005009

\bibitem[\protect\citeauthoryear{Brown}{1990}]{brown90} Brown, T.~M.: 1990,
'High Precision Doppler Measurements via Echelle Spectroscopy',
in {\it ASP Conf. Ser. 8: CCDs in Astronomy}, G.~H. Jacoby (ed.), ASP,
San~Franciso, pp.~335--344

\bibitem[\protect\citeauthoryear{Butler et al.}{2002}]{butler02} 
Butler, R.~P., Marcy, G.~W., Vogt, S.~S., Fischer, D.~A., Henry, G.~W., 
Laughlin, G., and Wright, J.:  2002,  'Seven New Planets Orbiting
G \& K Dwarfs', {\it Astrophysical Journal} (submitted)

\bibitem[\protect\citeauthoryear{Butler et al.}{2001}]{butler01} 
Butler, R.~P., et al.: 2001  'Statistical Properties of Extrasolar
Planets', in {\it Planetary Systems in the Universe: Observation, Formation, 
and Evolution}, A.~J. Penny, P. Artymowicz, A.-M. Lagrange, and
S. S. Russell (eds.), ASP, San Francisco

\bibitem[\protect\citeauthoryear{Butler et al.}{2000}]{butler00} 
Butler, R.~P., Vogt, S.~S., Marcy, G.~W., Fischer, D.~A., Henry, G.~W., and 
Apps, K.:  2000,  'Planetary Companions to the Metal-rich Stars BD~-10{\deg}3166 
and HD~52265', {\it Astrophysical Journal} {\bf 545}, pp.~504--511

\bibitem[\protect\citeauthoryear{Castellano}{2000}]{castellano00} 
Castellano, T.:  2000,  'A Search for Planetary Transits of the Star HD~187123 
by Spot Filter CCD Differential Photometry', {\it Publications of 
the Astronomical Society of the Pacific} {\bf 112}, pp.~821--826  

\bibitem[\protect\citeauthoryear{Charbonneau}{2003}]{charbonneau03} 
Charbonneau, D.: 2003, 'HD~209458 and the Power of the Dark Side',
in {\it ASP Conf. Ser.: Scientific Frontiers in Research on Extrasolar Planets}, D. Deming 
and S. Seager (eds.), ASP, San Francisco, astro-ph/0209517

\bibitem[\protect\citeauthoryear{Charbonneau et 
al.}{2000}]{charbonneau00} Charbonneau, D., Brown, T.~M., Latham, 
D.~W., and Mayor, M.:  2000,  'Detection of Planetary Transits Across a 
Sun-like Star', {\it Astrophysical Journal} {\bf 529}, 
pp.~L45--L48

\bibitem[\protect\citeauthoryear{Dreizler et 
al.}{2002}]{dreizler02} Dreizler, S., Rauch, T., Hauschildt, P., 
Schuh, S.~L., Kley, W., and Werner, K.:  2002,  'Spectral Types of 
Planetary Host Star Candidates: Two New Transiting Planets?', {\it 
Astronomy and Astrophysics} {\bf 391}, pp.~L17--L20

\bibitem[\protect\citeauthoryear{Fischer et 
al.}{1999}]{fischer99} Fischer, D.~A., Marcy, G.~W., Butler, 
R.~P., Vogt, S.~S., and Apps, K.:  1999,  'Planetary Companions around Two 
Solar-type Stars: HD~195019 and HD~217107', {\it Publications of the 
Astronomical Society of the Pacific} {\bf 111}, pp.~50--56  

\bibitem[\protect\citeauthoryear{Giampapa et 
al.}{1995}]{giampapa95} Giampapa, M.~S., Craine, E.~R., and Hott, 
D.~A.:  1995,  'Comments on the Photometric Method for the Detection of 
Extrasolar Planets.', {\it Icarus} {\bf 118}, pp.~199--210

\bibitem[\protect\citeauthoryear{Gilliland et 
al.}{2000}]{gilliland00} Gilliland, R.~L.~and 23 colleagues:  2000, 
'A Lack of Planets in 47 Tucanae from a Hubble Space Telescope Search', 
{\it Astrophysical Journal} {\bf 545}, pp.~L47--L51  

\bibitem[\protect\citeauthoryear{Henry et al.}{2002}]{henry02} 
Henry, G.~W., Donahue, R.~A., and Baliunas, S.~L.:  2002,  'A False Planet 
around HD~192263', {\it Astrophysical Journal} {\bf 577}, 
pp.~L111--L114  

\bibitem[\protect\citeauthoryear{Henry}{2000}]{henry00} Henry, 
G.~W.:  2000,  'Search for Transits of a Short-period, Sub-saturn 
Extrasolar Planet Orbiting HD~46375', {\it Astrophysical Journal} {\bf 
536}, pp.~L47--L48  

\bibitem[\protect\citeauthoryear{Henry et al.}{2000}]{henry00a} 
Henry, G.~W., Baliunas, S.~L., Donahue, R.~A., Fekel, F.~C., and Soon, W.:  
2000,  'Photometric and Ca~II~H and K Spectroscopic Variations in Nearby 
Sun-like Stars with Planets.~III.', {\it Astrophysical Journal} {\bf 
531}, pp.~415--437  

\bibitem[\protect\citeauthoryear{Henry et al.}{1997}]{henry97} 
Henry, G.~W., Baliunas, S.~L., Donahue, R.~A., Soon, W.~H., and Saar, 
S.~H.:  1997,  'Properties of Sun-like Stars with Planets: 51~Pegasi, 
47~Ursae~Majoris, 70~Virginis, and HD~114762', {\it Astrophysical Journal} 
{\bf 474}, pp.~503

\bibitem[\protect\citeauthoryear{Horne}{2003}]{horne03} 
Horne, K.: 2003, 'Status and Prospects of Transit Searches: Hot Jupiters Galore',
in {\it ASP Conf. Ser.: Scientific Frontiers in Research on Extrasolar Planets}, D. Deming 
and S. Seager (eds.), ASP, San Francisco

\bibitem[\protect\citeauthoryear{Janes}{1996}]{janes96} Janes, 
K.:  1996,  'Star Clusters: Optimal Targets for a Photometric Planetary 
Search Program', {\it Journal Geophysical Research} {\bf 101}, 
pp.~14853--14860  

\bibitem[\protect\citeauthoryear{Konacki et al.}{2003}]{konacki03} Konacki, 
M., Torres, G., Jha, S., and Sasselov, D.~D.:  2003,  
'A New Transiting Extrasolar Giant Planet', {\it Nature} 
(in press), astro-ph/0301052

\bibitem[\protect\citeauthoryear{Mallen-Ornelas et al.}{2002}]{mallen02}
Mallen-Ornelas, G., Seager, S., Yee, H.~K.~C., Minniti, D., Gladders, M.~D., 
Mallen-Fullerton, G., and Brown, T.~M.: 2002, 'The EXPLORE Project I:
A Deep Search for Transiting Extrasolar Planets', {\it Astrophysical Journal} (in press),
astro-ph/0203218

\bibitem[\protect\citeauthoryear{Marcy et al.}{2000}]{marcy00} Marcy, G.~W.,
Cochran, W.~D., and Mayor, M.: 2000, 'Extrasolar Planets around Main-sequence Stars', 
in {\it Protostars and Planets IV},
V. Mannings, A.~P. Boss, and S.~S. Russell (eds.), Univ. of Arizona Press, Tucson,
pp.~1285

\bibitem[\protect\citeauthoryear{Mayor and 
Queloz}{1995}]{mayor95} Mayor, M.~and Queloz, D.:  1995,  'A 
Jupiter-mass Companion to a Solar-type Star', {\it Nature} {\bf 
378}, pp.~355

\bibitem[\protect\citeauthoryear{Mochejska et 
al.}{2002}]{mochejska02} Mochejska, B.~J., Stanek, K.~Z., Sasselov, 
D.~D., and Szentgyorgyi, A.~H.:  2002,  'Planets in Stellar Clusters 
Extensive Search. I. Discovery of 47 Low-amplitude Variables in the 
Metal-rich Cluster NGC 6791 with Millimagnitude Image Subtraction 
Photometry', {\it Astronomical Journal} {\bf 123}, pp.~3460--3472

\bibitem[\protect\citeauthoryear{Pepe et al.}{2002}]{pepe02} 
Pepe, F., Mayor, M., Galland, F., Naef, D., Queloz, D., Santos, N.~C., 
Udry, S., and Burnet, M.:  2002,  'The CORALIE Survey for Southern 
Extra-solar Planets VII. Two Short-period Saturnian Companions to 
HD~108147 and HD~168746', {\it Astronomy and Astrophysics} {\bf 388}, 
pp.~632--638  

\bibitem[\protect\citeauthoryear{Rosenblatt}{1971}]{rosenblatt71} 
Rosenblatt, F.:  1971,  'A Two-color Photometric Method for Detection of 
Extra Solar Planetary Systems', {\it Icarus} {\bf 14}, pp.~71

\bibitem[\protect\citeauthoryear{Santos et al.}{2000}]{santos00} 
Santos, N.~C., Mayor, M., Naef, D., Pepe, F., Queloz, D., Udry, S., Burnet, 
M., and Revaz, Y.:  2000,  'The CORALIE Survey for Southern Extra-solar 
Planets. III. A Giant Planet in Orbit around HD~192263', {\it Astronomy and 
Astrophysics} {\bf 356}, pp.~599--602  

\bibitem[\protect\citeauthoryear{Street et al.}{2002}]{street02} 
Street, R.~A.~and 9 colleagues:  2002,  'Variable Stars in the Field of 
Open Cluster NGC~6819', {\it Monthly Notices of the Royal Astronomical 
Society} {\bf 330}, pp.~737--754

\bibitem[\protect\citeauthoryear{Tinney et al.}{2001}]{tinney01} 
Tinney, C.~G., Butler, R.~P., Marcy, G.~W., Jones, H.~R.~A., Penny, A.~J., 
Vogt, S.~S., Apps, K., and Henry, G.~W.:  2001,  'First Results from the 
Anglo-Australian Planet Search: A Brown Dwarf Candidate and a 51~Peg-like 
Planet', {\it Astrophysical Journal} {\bf 551}, pp.~507--511  

\bibitem[\protect\citeauthoryear{Udalski et 
al.}{2002a}]{udalski02} Udalski, A.~and 8 colleagues:  2002,  'The 
Optical Gravitational Lensing Experiment. Search for Planetary and 
Low-luminosity Object Transits in the Galactic Disk. Results of 2001 
Campaign', {\it Acta Astronomica} {\bf 52}, pp.~1--37

\bibitem[\protect\citeauthoryear{Udalski et 
al.}{2002b}]{udalski02a} Udalski, A., Zebrun, K., Szymanski, M., 
Kubiak, M., Soszynski, I., Szewczyk, O., Wyrzykowski, L., and Pietrzynski, 
G.:  2002,  'The Optical Gravitational Lensing Experiment. Search for 
Planetary and Low-luminosity Object Transits in the Galactic Disk. Results 
of 2001 Campaign -- Supplement', {\it Acta Astronomica} {\bf 52}, 
pp.~115--128

\bibitem[\protect\citeauthoryear{Vogt et al.}{2000}]{vogt00} 
Vogt, S.~S., Marcy, G.~W., Butler, R.~P., and Apps, K.:  2000,  'Six New 
Planets from the Keck Precision Velocity Survey', {\it Astrophysical 
Journal} {\bf 536}, pp.~902--914  

\bibitem[\protect\citeauthoryear{Yee et al.}{2002}]{yee02}
Yee, H.~K.~C., et al.: 2002, 'The EXPLORE Project:
A Deep Search for Transiting Extra-solar Planets', in
{\it Proceedings of the SPIE Conference: Astronomical Telescopes and Instrumentation}
(in press), astro-ph/0208355

\end{thebibliography}
\end{document}